# Silica final lens performance in laser fusion facilities: HiPER and LIFE

**D Garoz[1], R González-Arrabal[1], R Juárez[1,2], J Álvarez[1], J Sanz[1,2], J M Perlado[1] and A Rivera[1]**

[1]Instituto de Fusión Nuclear, UPM, Jose Gutierrez Abascal 2, E28006 Madrid, Spain.
[2]Departamento de ingeniería energética, UNED, Juan del Rosal 12, E28040 Madrid, Spain.

E-mail: david.garoz@upm.es

**Abstract.** Nowadays, the projects LIFE (Laser Inertial Fusion Energy) in USA and HiPER (High Power Laser Energy Research) in Europe are the most advanced ones to demonstrate laser fusion energy viability. One of the main points of concern to properly achieve ignition is the performance of the final optics (lenses) under the severe irradiation conditions that take place in fusion facilities. In this paper, we calculate the radiation fluxes and doses as well as the radiation-induced temperature enhancement and colour centre formation in final lenses assuming realistic geometrical configurations for HiPER and LIFE. On these bases, the mechanical stresses generated by the established temperature gradients are evaluated showing that from a mechanical point of view lenses only fulfill specifications if ions resulting from the imploding target are mitigated. The absorption coefficient of the lenses is calculated during reactor startup and steady-state operation. The obtained results evidence the necessity of new solutions to tackle ignition problems during the startup process for HiPER. Finally, we evaluated the effect of temperature gradients on focal length changes and lens surface deformations. In summary, we discuss the capabilities and weak points of silica lenses and propose alternatives to overcome predictable problems.

## 1. Introduction

Fusion energy is foreseen to become within the next two decades a real competitor of fossil fuels with the added advantage of small environmental impact, safety and minimal waste generation. Nowadays, with the National Ignition Facility (NIF) in the last optimization stages to demonstrate ignition with energy gain, there is an increasing interest in nuclear fusion by inertial confinement with laser (laser fusion) as a commercial power source. Currently, the projects LIFE (Laser Inertial Fusion Energy) in USA and HiPER (High Power Laser Energy Research) in Europe are the most advanced ones to demonstrate laser fusion energy viability with indirect and direct drive targets, respectively.

The HiPER project is at the end of the preparatory phase (phase 2). The next step (phase 3), will take place over the next 7 years to carry out appropriate R&D activities that eventually will lead to the construction of a demonstration (demo) power plant (phase 4). Since HiPER has adopted a risk reduction strategy based on achieving milestones before moving from one phase to the next, phase 4 indeed can be divided into two sub-phases: Phase 4a, construction of an experimental facility to demonstrate an advanced ignition scheme and repetitive laser operation; Phase 4b, construction of a demo power plant. Furthermore, we can devise a prototype facility as an intermediate step between the experimental facility and the demo power plant. The prototype facility does not imply any further construction work but a relaxed operation mode. This can be accomplished by using low yield targets that allow one to study aspects like target injection, tracking, repetition mode, heat extraction or tritium production while keeping materials demands low. Moreover, the prototype facility operation mode will be especially useful to test materials under irradiation (as devised in LIFE.1 [1], described next). Based on the NIF construction experience, the LIFE project [1, 2, 3, 4] aims at the construction of a demo power plant, first with available technologies and existing materials (LIFE.1), and subsequently with improved capabilities (LIFE.2) based on the development of new technologies and materials in LIFE.1. A brief overview of the different development stages for HiPER project and LIFE is summarized in table 1.

Table 1. HiPER development strategy and LIFE scenarios.

| | Experimental facility | Prototype plant | Demo plant | LIFE.1 | LIFE.2 |
|---|---|---|---|---|---|
| Operation | Bunches of 100 shots, max. 5 DT explosion | Continuous (24/7) | Continuous (24/7) | Continuous (24/7) | Continuous (24/7) |
| Yield (MJ) | <20 | <50 | >100 | 27 | 132 |
| Rep. rate (Hz) | 1-10 | 1-10 | 10-20 | 16 | 16 |
| Power (GWt) | - | < 0.5 | 1-3 | 0.4 | 2.2 |
| T cycle | No | Yes | Yes | Yes | Yes |
| Blanket | No | Yes | Yes | Yes | Yes |

The development of materials able to withstand the harsh reactor environment is one of the main challenges to make fusion energy a reality. In particular, the development of materials with improved properties for the final optics components is a main point of concern since the ignition process itself depends on them. A special concern relies on the final lenses because they must face the target explosions only a few meters away during operation. Appropriate lenses must present (i) low laser absorption, (ii) good thermo-mechanical properties and (iii) high radiation resistance. Nowadays silica is proposed as the best candidate for final lenses, because of its good optical transparency from around 300 nm to the visible band (covering the $2^{nd}$ and $3^{rd}$ harmonic wavelengths of a typical solid-state laser), good thermo-mechanical properties, high radiation resistance and low cost [5, 6].

Previous works evidence that radiation generates point defects in silica. It is well known that fast neutron irradiation of silica gives rise to two main effects: network compaction (densification) [7] and point defect generation [8, 9]. On the other hand, purely ionizing radiation plays a role in the formation of colour centres [10, 11, 12] which originates undesired laser absorption and scattering [13]. Moreover, simultaneous neutron and γ-ray irradiation (as it is the case in a laser fusion reactor) leads to synergistic effects resulting in an enhanced degradation of the optical transmission properties [6, 14, 15]. In general, defects in silica are very complex. They can be charged or uncharged and interact among themselves giving rise to different configurations depending on temperature, stoichiometry and irradiation conditions [16, 17]. Moreover, extrinsic defects, which can be inherent to the fabrication process, in particular hydrogen-related defects, play a crucial role in the radiation-induced defect configuration [18].

The purpose of this work is to estimate the performance of silica final lenses under realistic irradiation conditions for HiPER and LIFE.2 facilities (table 1). For this purpose, firstly the radiation fluxes in the lenses are estimated. Secondly, the radiation induced temperature enhancement, stress generation and colour centre formation (at a constant temperature) are determined. From these results the thermo-mechanical and optical (absorption) properties of the lenses are evaluated. Finally, the temperature enhancement and colour centre evolution during reactor startup are calculated. In summary, we illustrate the capabilities and weak points of silica lenses as final lenses and propose alternatives to overcome predictable problems.

## 2. Chamber design

HiPER chamber design is currently underway. However, some advanced concepts have been already studied [19, 20]. It has been proposed a spherical reactor chamber for the experimental facility with a 5 m inner radius and 48 openings for the laser beam lines for symmetrical target illumination. The 10 cm thick chamber walls are surrounded by borated concrete to stop neutrons. On the other hand, the demo and prototype facilities have been devised as a 6.5 m inner radius steel chamber surrounded by a 75 cm thick liquid metal blanket (figure 1). In all cases lenses are located 8 m away from the chamber centre, to assure a good focal spot on the target. As indicated in figure 1, a thick concrete wall is located 16 m away from the chamber centre to protect the final optics components (except the final lenses). The laser beam passes through the wall by a thin opening (pinhole). The dimensions of the silica lenses are 75 x 75 cm$^2$ with a thickness of 5 cm. Due to the circular shape of the chamber openings, only a 60 cm diameter circular area is directly exposed to the target. LIFE [3] will use 0.5 cm-thick Fresnel final lenses located at 17 meter from the chamber center with dimensions large enough to accommodate the laser beams (48 x 48 cm$^2$).

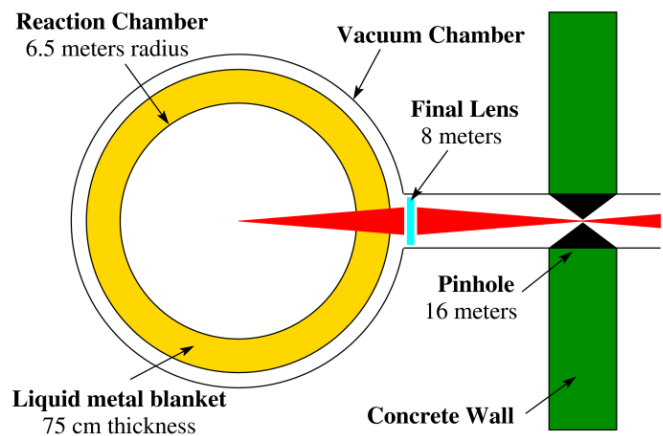

**Figure 1.** Schematic representation of the HiPER prototype and demo plants. Distances from the chamber centre are indicated.

## 3. Radiation fluxes

The radiation nature and fluxes in laser fusion depend on the type of target. In this work we follow the studies of direct and indirect targets carried out in the frame of the ARIES project [21]. Thus, we use the ARIES direct drive target with a yield of 154 MJ to study the irradiation conditions of the final lenses in a HiPER demo power plant. For the prototype facility, the radiation spectra are rescaled to a target yield of 50 MJ (table 1). In the case of a direct drive target, the most significant contributions are due to fusion neutrons (~71% of target yield) and ions (burn products and debris) that carry nearly 27% of the total energy released by the explosion [21, 22, 23]. In the case of indirect drive targets (as in LIFE) in addition to fusion neutrons (~69% of the target yield) and to high energy ions which carry ~6% of the total energy released by the explosion, a large fraction of X-rays are generated due to the hohlraum accounting up to 25% of the target yield [21, 22, 23]. Thus, the major difference is the important X-ray contribution generated in sub-ns timescales with indirect targets leading to a huge power deposition on the surrounding materials, which could not withstand the thermo-mechanical shock without appropriate mitigation strategies (for example, filling the chamber with residual gas) [4, 22, 24].

On this basis, the neutron and γ-ray doses absorbed by the silica lenses in HiPER were accurately calculated as a function of time with MCNPX [25] for the HiPER prototype and demo reactor geometry described in section 2. The reactor geometry was designed with CATIA and converted with MCAM [26] into a valid geometrical input for MCNPX. The mean free paths were obtained from the ENDF-VII data base [27]. The results for HiPER demo reactor are shown in figure 2. A high neutron flux (corresponding to primary neutrons) is observed immediately after the explosion which rapidly decreases up to three orders of magnitude within the first 60 ns. The total neutron dose rate calculated with MCNPX follows the same trend. Due to (n, γ) reactions a significant γ-pulse accompanies the emission of neutrons. These results indicate that the final lenses receive concomitant neutron and gamma pulses after every explosion. It is worthwhile to mention that for the prototype reactor the curves follow same time dependence but the absolute values are a factor 1/3 lower than those depicted in figure 2.

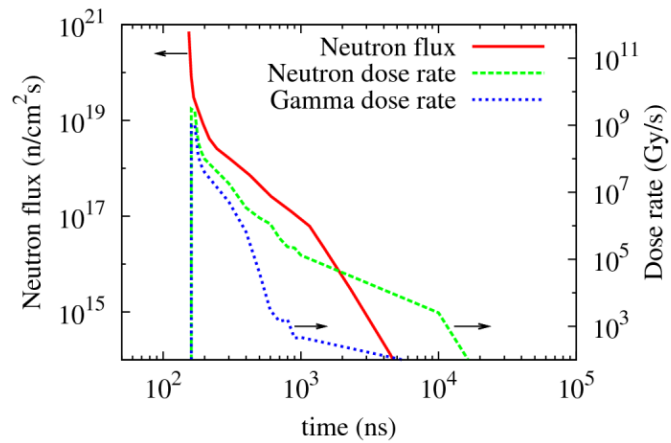

**Figure 2.** Primary neutron flux, total neutron and gamma dose rates as a function of time after each explosion.

The neutron and γ-ray doses absorbed by the final lenses in LIFE.2 were estimated from the 458 MJ HI indirect-drive target [21] rescaled to 132 MJ. Because of the unknown complete reactor chamber geometry, the neutron flux ($\Phi_n$) was calculated using only direct neutrons of energy $E_n$. Then, the neutron dose rate was obtained as follows

$$\dot{D}_n = \frac{\Phi_n E_{PKA}}{d \rho_{SiO2}} \left[ 1 - \exp\left(\frac{-d}{\lambda_n}\right) \right] \text{ with } E_{PKA} = \frac{2}{M + 2/3} E_n , \qquad (1)$$

where $\lambda_n$ is the mean free path, $d$ the lens thickness, $\rho_{SiO2}$ the density and $M$ the average atomic mass (20 for $SiO_2$). According to the previous calculations carried out with MCNPX for the HiPER demo reactor (figure 2) and for the HiPER experimental facility [28] the gamma dose is assumed to be 0.4 times the neutron dose.

The estimated radiation average energy, pulse width and mean range (penetration depth) values at HiPER final lenses are depicted in table 2. Mean energy and pulse width have been estimated from ARIES direct drive with a yield of 154 MJ [21, 29], except X-ray pulse width taken from reference [30]. Ion mean range have been calculated by means of SRIM code [31] and X-ray mean range using appropriated absorption coefficients [32]. In addition, the total energy density (ED) deposited by each radiation form is also shown for HiPER facilities and LIFE.2 lenses, which are located at 8 m and 17 m respectively. As shown in table 2, the deposited energy by neutrons and gammas in LIFE.2 and in HiPER prototype lenses are quite similar.

**Table 2.** Mean energy, pulse width, mean range and energy density (ED) deposited in lenses by each radiation form in different scenarios for HiPER and LIFE.2

|  | HiPER (8 m) | | | | | | LIFE.2 (17 m) |
|---|---|---|---|---|---|---|---|
|  | Mean Energy (MeV) | Pulse width (ns) | Mean range (μm) | ED in experim. (J/cm³) | ED prototype (J/cm³) | ED demo (J/cm³) | ED LIFE.2 (J/cm³) |
| Burn products (⁴He) | 2.1 | 400 | 6.4 | 492 | 1230 | 3788 | -[a] |
| Debris ions (D,T) | 0.15 | 2200 | 1.4 | 2549 | 6372.7 | 19628 | -[a] |
| X-rays | 0.007 | 0.17 | 10 | 34 | 85 | 261 | -[a] |
| Neutrons | 12.4 | 60 | - | 0.018 | 0.046 | 0.142 | 0.03 |
| Gammas | - | ≈60 | - | 0.007 | 0.017 | 0.051 | 0.012 |

[a] Gas protection avoids thermal shock due to ions and X-rays in LIFE lenses.

## 4. Thermo-mechanical response

The radiation-induced thermo-mechanical response of the lenses was calculated by means of the finite element solver Code Aster [33]. In order to speed up calculations, the lenses were assumed to have cylindrical geometry with a diameter of 75 cm and a thickness of 5 cm for HiPER and a diameter of 70 cm and thickness of 0.5 cm for LIFE.2. As shown in figure 3, these geometries were modeled by an axis-symmetric 2D mesh using Salome Platform [34]. Moreover, to achieve a detailed estimation of the temperature gradients and local stresses, the mesh is refined with small elements of 100 nm along axial direction $(z)$ at the inner surface and in radial direction $(r)$ at the irradiated/unirradiated boundary ($r=30$ cm for HiPER and $r= 24$ cm for LIFE.2). The lens surfaces were supposed to emit radiation and the lens surrounding temperature was considered to be constant. The radiation emitted through the lens surfaces was calculated according to the Stephan-Boltzmann law. The mechanical boundary conditions are determined by the axis-symmetric geometry.

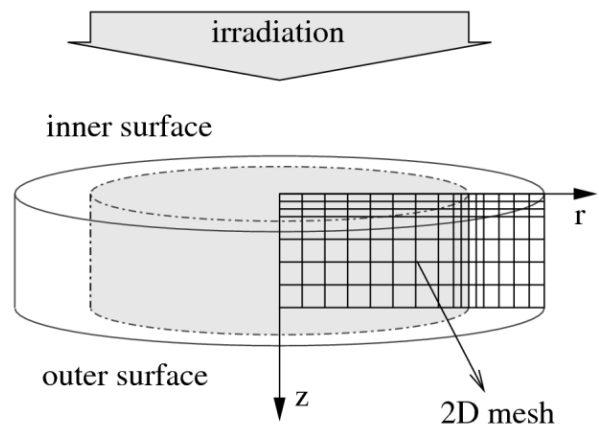

**Figure 3.** Representation of a cylindrical lens irradiated parallel to the lens axis ($z$). The 2D mesh employed for the calculations is schematically shown. It is formed by elements of different dimensions.

*4.1. HiPER*

After every explosion the final lenses have to withstand high thermal loads in the form of ions, X-rays, neutrons and gammas. The radiation induced thermal loads mainly depend on the radiation flux and nature. Since the irradiation in laser fusion is pulsed (consequence of target explosions), the energy deposition takes place within a few microseconds after the explosion, starting with X-ray and finishing with the last debris ions that reach the lens (table 2). Ions and X-rays deposit almost all their energy in the first few microns (<10 μm) beneath the inner lens surface, whereas neutrons and gammas deliver their energy (almost) homogeneously along the lens volume.

As shown in table 2, the energy density deposited by ions is 1230 J/cm$^3$ and 3788 J/cm$^3$ for HiPER prototype and demo reactors, respectively. Such a density would lead to temperature enhancements higher than the silica melting temperature in just one shot. Therefore, the ions must be somehow mitigated [35]. Details on ion mitigation are beyond the scope of this paper. We will simply assume from now on that ion mitigation occurs.

The experimental facility will operate in bunch mode at room temperature, therefore, the lens temperature prior to the pulse arrival can be considered uniform. On the other hand, the prototype and demo reactors will operate in continuous mode. Assuming for the prototype reactor a steady state situation the lens surface temperature will be 866 K and higher in the centre at the beginning of the pulse (see discussion below and figure 5(a)). Disregarding ions, the major contribution to the temperature enhancement at depths ≤ 10 µm is due to X-rays. The radiation-induced temperature enhancement as a function of time after one explosion at several depths underneath the lens inner surface is depicted in figures 4(a) and 4(b) for the HiPER experimental and prototype reactor, respectively. The surface temperature increases 7 K for the experimental and 15 K for the prototype reactor. The temperature drops fast as a function of time after the explosion and as a function of the distance from the lens surface. The temperature gradient at depths ≤ 10 µm disappear after 100 µs and the temperature at a depth of 10 µm increases only 2 K due to the heat transferred by conduction from the inner surface. In the case of the prototype reactor, the neutron and γ-ray contribution leads to a temperature rise of about 0.1 K per shot along the whole lens thickness. This contribution is negligible for the experimental reactor. The temperature rise generates cyclic stress at the irradiated inner surface depths ≤ 10 µm, see figures 4(c) and 4(d). The X-ray thermal shock increases the volume of the inner surface material generating compression stress, In the case of the prototype facility, figure 4(d), the initial traction radial stress due to the initial temperature decreases until compression values are reached at layers ≤ 0.5 µm.

When working in continuous mode, the average lens temperature increases if the energy deposited in one pulse is higher than that radiated by the lens surfaces. When both contributions balance each other, steady state is achieved. Assuming that the lens temperature equals the surrounding temperature ($T_o$ = 600 K) before startup reactor, the steady state maximum temperature into the lens reaches 938 K for the prototype reactor and 1304 K for demo reactor, below and above the maximum silica service temperature (1223 K), respectively, see table 3. The steady state situation occurs after 32000 pulses for the prototype reactor and after 25000 pulses for the demo reactor. Note that in continuous mode the neutron and gamma contributions are very relevant since (ignoring ions) they carry most of the energy. The energy deposited by laser absorption is negligible (and therefore not considered here) because the high temperature reached in continuous mode keeps the optical absorption low (see section 5). Therefore, even when disregarding ions, we conclude that silica lenses cannot operate under HiPER demo reactor conditions in the present configuration. A possible solution could be to use external coolers for the lenses or modify the configuration moving the final lenses further away from the chamber centre. Both possibilities imply a detailed study beyond the scope of this work.

The 2D temperature profile and stresses in steady state conditions for the prototype facility are depicted in figure 5(a). The lenses reach a maximum internal temperature of 938 K, with 866 K at the inner surface where X-rays are deposited and 826 K at the outer surface. The coolest area is the lateral surface ($r$ = 37.5 cm) which corresponds to the unirradiated volume. The temperature gradients along the $z$ and $r$ directions lead to stress generation. The temperature distribution in $z$ is mainly responsible for radial and azimuthal gradient stresses within the irradiated volume. In particular, tractions of 0.46 MPa and of 0.71 MPa are observed at the lens inner and outer surfaces, respectively. A compressive stress of -2.75 MPa is calculated at the centre of the lens where the temperature reaches its maximum value. Significant contribution neither to axial nor to shear stress are observed related to the temperature distribution along $z$. The fact that the temperature of the unirradiated volume is around 600 K impedes the lens expansion in $r$ directions. A compressive stress for $r$ < 30cm, traction for $r$ > 30 cm, as well as axial and shear stresses at $r$ = 30 cm are generated due to the temperature gradient along the $r$ direction. In all cases the calculated stresses are observed to be lower than the silica tensile strength (48 MPa) which indicates that, as a fair simplification, silica lenses can withstand the radiation-induced mechanical stresses.

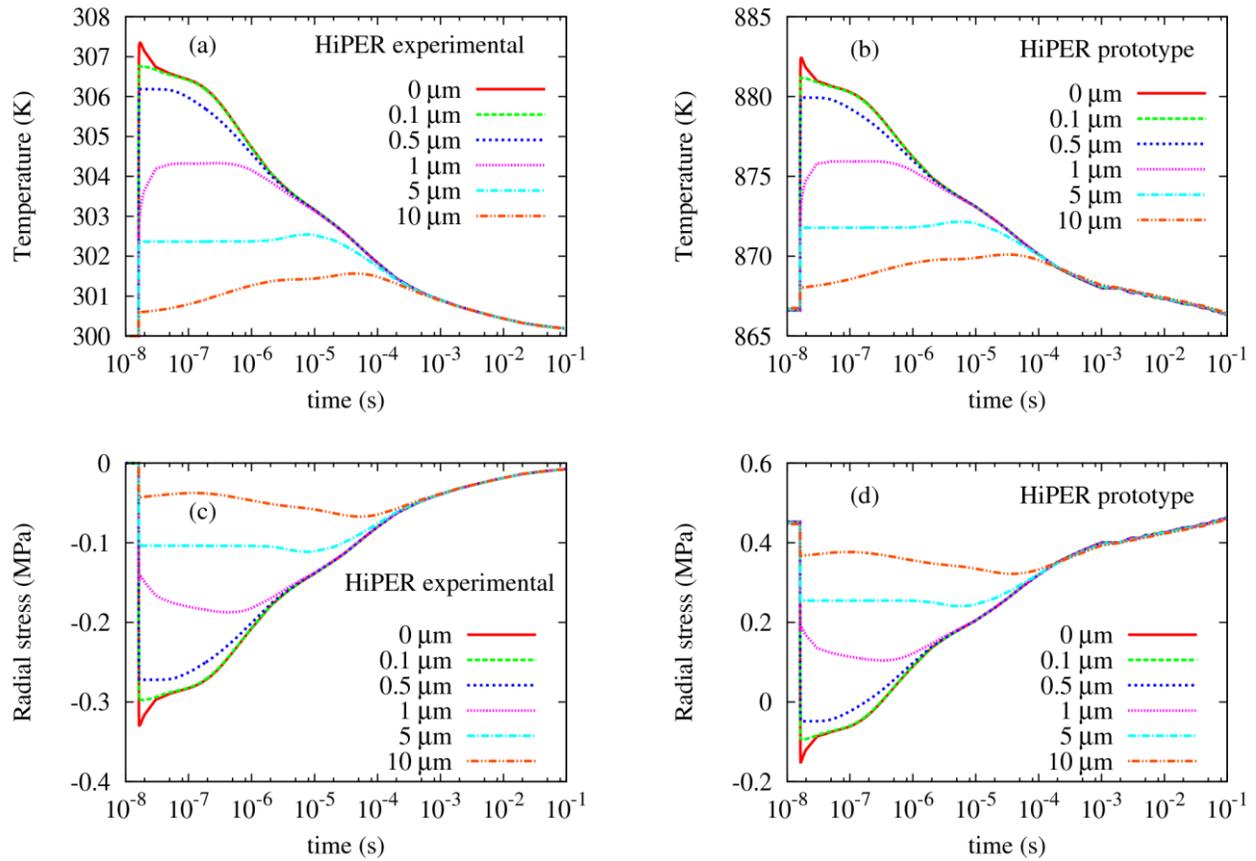

**Figure 4.** Final lens temperature as function of time after each explosion for different depths in the HiPER experimental facility (a) and in the prototype facility (b). Radial stress evolution at different depths for experimental facility (c) and for the prototype facility (d).

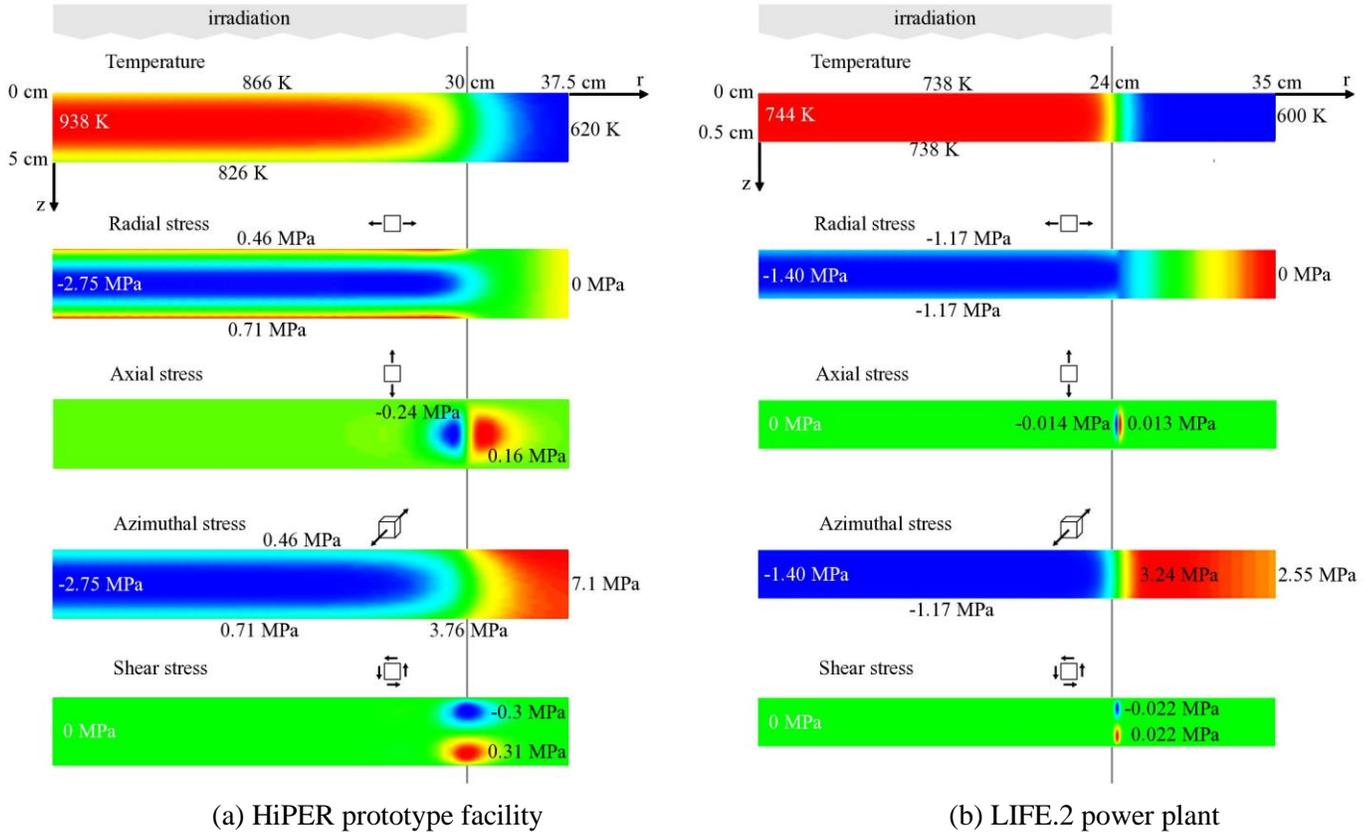

(a) HiPER prototype facility          (b) LIFE.2 power plant

**Figure 5.** 2D colour maps of temperature and of radial, axial, azimuthal and shear stresses at the end of each pulse in steady-state operations conditions for HiPER prototype facility (a) and for LIFE.2 facility (b).

In order to have a more detailed picture of silica mechanical behavior it must be considered that silica can suffer brittle fracture by crack generation and growth. The lenses can have preexisting cracks due to the fabrication process or due to their manipulation. In addition, operation may lead to crack generation stemming from irradiation induced-point defects. Moreover, the impact of shrapnel from the target explosions (especially relevant with indirect targets) may generate cracks at the inner surface of lenses.

The intensity factor in traction (mode I [36]), $K_I = \sigma_{max}\sqrt{\pi a}$, allows one to predict the stress state near the tip of an existing crack with a length of $2a$ caused by the maximum traction stress $\sigma_{max}$. Under the assumption of linear elastic fracture mechanics [36], cracks start to grow when the intensity factor is equal to or larger than the fracture toughness ($K_{IC}$ = 0.73 MPa/m$^{1/2}$ for silica [37]). Therefore, knowing the maximum traction stress for each scenario, one can estimate the critical crack length for brittle fracture to occur. As shown in table 3, the critical crack length for the HiPER experimental facility is higher than the lens dimensions, which indicates that silica lenses will not suffer brittle fracture. The estimated critical length for HiPER prototype and demo reactors are 6.73 mm and 2.04 mm, respectively. Thus, longer cracks would lead to brittle fracture of the lenses. In order to prevent the failure of HiPER lenses in continuous mode, preexisting cracks near the lateral surface ($r$ = 37.5 cm) must be avoided during manufacture and subsequent manipulation.

Moreover, one has also to consider the effect of fatigue. Cracks can grow due to cyclic fatigue induced by traction cyclic stresses. The X-ray thermal shock generates cyclic stresses near the inner surface. In the experimental facility the axial traction stress at $r$ = 30 cm varies from 0.12 MPa to zero along a distance of around 1 µm (not shown in figures). The cyclic stresses in prototype and demo extend up to 1 µm depth at the irradiated inner surface, see figure 4(d) for prototype. The traction stress range $\Delta\sigma$ is 0.12 MPa for the experimental facility, 0.46 MPa for the prototype reactor and 0.52 MPa for the demo reactor. We can estimate the stress intensity factor range for each facility with $\Delta K_I = \Delta\sigma\sqrt{\pi a}$, assuming 1 µm crack length, which is equal to the volume affected by the traction cyclic stresses. The stress intensity factor ranges have been estimated in table 3. In all scenarios the stress intensity factor range is well below the fatigue threshold of silica, $\Delta K_{th}$ = 0.3 MPa/m$^{1/2}$ [38], therefore the probability of crack growth due to cyclic fatigue is negligible.

In conclusion, silica lenses in the experimental and prototype facility appear to withstand the thermo-mechanical conditions imposed by the irradiation conditions (and assuming ion mitigation). On the other hand, the lenses will not operate under the demo reactor extreme irradiation conditions (even assuming ion mitigation).

**Table 3.** Calculated thermo-mechanical values and critical limits for silica. $T_{max}$ is the maximum temperature, $\sigma_{max}$ the maximum traction stress, $(2a)_c$ the critical crack length and $\Delta K_I$ the stress intensity factor range assuming 1 µm-long cracks.

|  | HiPER experimental | HiPER prototype | HiPER demo | LIFE.2 | Critical limits of silica |
|---|---|---|---|---|---|
| $T_{max}$ (K) | 620 | 938 | 1300 | 660 | 1223 |
| $\sigma_{max}$ (MPa) | 0.12 | 7.1 | 9.2 | 3.25 | 48 |
| $(2a)_c$ (mm) | >2×10$^4$ | 6.73 | 2.04 | 32.12 | - |
| $\Delta K_I$ (Mpa/m$^{1/2}$) | 1.5×10$^{-4}$ | 5.8×10$^{-4}$ | 10$^{-3}$ | 0 | 0.3 |

*4.2 LIFE.2*

Due to the intense X-ray pulses resulting from indirect target explosions, mitigation strategies are a must to protect the chamber components. Usually, gas protection scenarios are proposed as it is the case for LIFE. Thus, the prompt irradiation that reaches the lenses is mainly in the form of neutron and gamma pulses. In order to properly calculate the energy deposited in the lenses one must also take into account the laser contribution because in this case the maximum temperatures are too low to keep the optical absorption low (see section 5). The heat flux due to laser irradiation is determined as follows [15],

$$q_{laser} = F_{laser}\nu(1-\exp(-Ad)), \qquad (2)$$

where $F_{laser}$ is the laser fluence, $\nu$ the frequency, $A$ the absorption coefficient and $d$ the lens thickness. By considering an optical absorption coefficient of 3.9% (see section 5) the heat flux is estimated to be 0.17 J/cm$^3$. On this basis, the total deposited heat energy density is calculated to be 0.212 J/cm$^3$. Such an energy deposition leads to a temperature rise of 0.01 K per pulse, which is almost homogeneous along the whole lens thickness. Assuming a surrounding temperature of 600 K, the pulsed steady state solution is reached after 18000 pulses, which results in a steady state maximum temperature of 744 K.

The 2D temperature profile and stresses in steady state conditions for LIFE.2 are represented in figure 5(b). Because of the homogeneous heat deposition along $z$ and the small lens thickness, the temperature profile along this direction varies only 6 K. Concerning to the $r$ axis, the temperature reaches a maximum of 744 K at $r = 0$ and decreases down to 600 K at $r = 35$ cm. This temperature gradient along $r$ drives to the generation of compression radial and azimuthal stresses of -1.4 MPa within the irradiated volume and a maximum azimuthal traction stress of 3.25 MPa at the lateral surface ($r = 35$ cm). In all cases, the calculated stresses are lower than those for HiPER prototype lenses and below the critical limit for silica (see table 3). In addition, the lenses of LIFE.2 can withstand larger cracks (32.12 mm) than the lenses of HiPER facilities. Since, this value is higher than the thickness of lenses, the LIFE Fresnel lenses are not expected to suffer from brittle fracture. In conclusion, the LIFE Fresnel lenses (located far from the explosions) can in principle withstand the radiation-induced thermo-mechanical loads. Nevertheless, it is important to remark that in these studies the effect of the surrounding gas on lenses has not been considered, which has to be done in order achieve a more precise picture.

**5. Colour centre formation**

As previously mentioned, irradiation induces point defect generation in silica. In this section the degradation of the lenses due to atomistic effects is studied by using a model developed by Marshall et al. [14] to account for the colour centre formation [14, 15, 28]. The model considers that neutron collisions with the silica network produce oxygen deficient centres (ODC) that can be radiolytically converted into E' centres via gamma interaction. Moreover, both ODC and E' centres can be annealed out if the temperature is high enough. A schematic view of the colour centre formation and annihilation is depicted in figure 6.

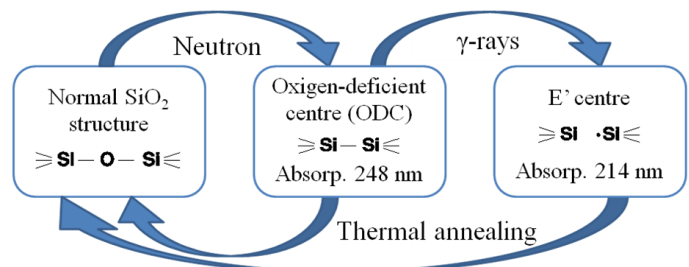

**Figure 6.** Schematic sketch of the colour centre formation and annihilation according to the model initially developed by Marshall et al [14].

The colour centre evolution is estimated for the HiPER prototype facility by assuming that the lenses are simultaneously subject to neutron and gamma irradiation (section 3) and their temperature is constant at 850 K. X-ray and ion fluxes are ignored. The calculated ODC and E' concentrations accumulated at two different repetition rates are shown in figure 7. It is observed that both ODC and E' centres are created in the first nanosecond after each explosion, see figure 7(a). Defect annealing takes place effectively at this temperature. In fact, defect saturation becomes evident beyond 1000 pulses at a repetition rate of 10 Hz and occurs after 150 pulses at 1 Hz. The ODC concentration is always higher than that for E' centres. This indicates that the γ-dose rate is too low to promote complete ODC conversion to E' centers.

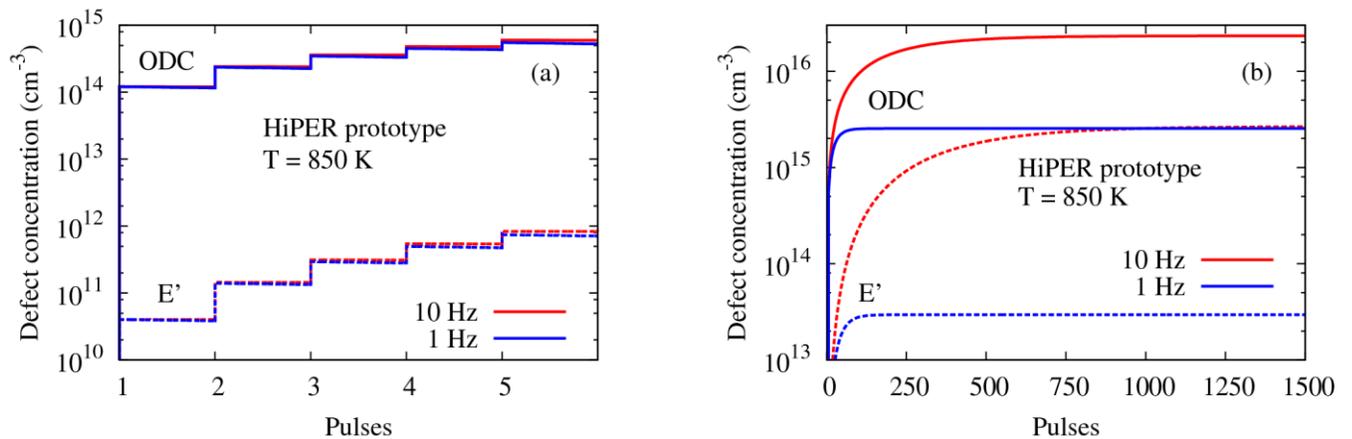

**Figure 7.** Concentration of oxygen defect centres (ODC) and E' centres accumulated in the final lens as a function of 50 MJ yield target explosions for two different repetition rates and constant lens temperature (annealing temperature) of 850 K.

The absorption coefficient of the lenses ($\alpha$) is related to the defect concentrations by the following equation

$$\alpha(\lambda) = \sum_i \sigma_i N_i L_i(\lambda) \text{ with } L_i(\lambda) = \frac{1}{1+\frac{(\lambda_i - \lambda)^2}{(\Delta\lambda_i)^2}}, \quad (3)$$

where the subscript $i$ denotes a defect center, $\sigma_i$ the defect center cross section, $N_i$ defect concentration, $\lambda$ the wavelength, $\lambda_i$ the peak wavelength for the defect center $i$ and $\Delta\lambda_i$ the half width at half maximum of the wavelenght for the centre $i$. The absorption coefficients calculated at a fixed temperature (850K) as a function of wavelengths for the HiPER prototype facility are shown in figure 8. The parameters used for the calculations are listed in table 4. Two peaks located at 248 nm and 214 nm, corresponding to ODC and E' centres, respectively, are clearly observed after the first 100 pulses. The intensity of both peaks notably grows with increasing pulse number until saturation (due to defect annealing) occurs for pulse number higher than 1000. This fact limits the lens absorption.

**Table 4.** Values used for the parameters of the model

| Parameter | Value | Ref. |
|---|---|---|
| $\sigma_{ODC}$ (cm$^2$) | $1.7\times10^{-17}$ | 15 |
| $\lambda_{ODC}$ (nm) | 248 | 39 |
| $\Delta\lambda_{ODC}$ (nm) | 13 | 39 |
| $\sigma_{E'}$ (cm$^2$) | $3.2\times10^{-17}$ | 40 |
| $\lambda_{E'}$ (nm) | 214 | 39 |
| $\Delta\lambda_{E'}$ (nm) | 15 | 39 |

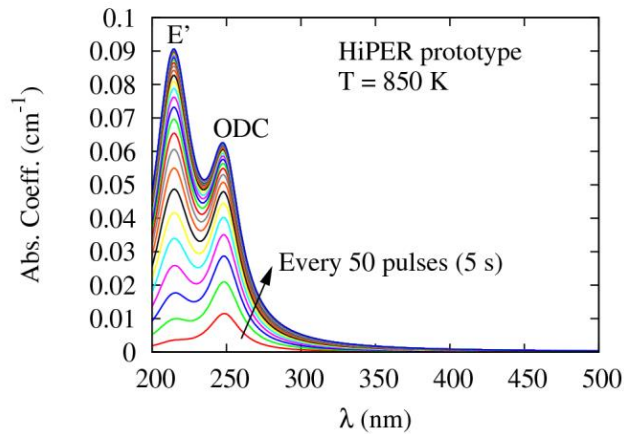

**Figure 8.** Absorption coefficient for a lens kept at 850 K during exposure to 50 MJ target explosion at a rate of 10 Hz (prototype facility conditions). The lowest curve correspond to 50 pulses, and the highest to 1500 (150 s) (the curves are given every 50 pulses, 5 s).

As previously mentioned one of the main parameters of the lenses is the optical transparency in the 350 nm region. Because of this reason, the evolution of the lens absorption for different temperatures is studied. The optical absorption, $A$, is calculated by means of

$$A = 1 - \exp(-\alpha(\lambda)d), \tag{4}$$

where $d$ is the lens thickness. The results for HiPER experimental and prototype facilities are shown in figure 9. For the experimental facility, the optical absorption reaches 5% in about 2000 pulses for temperatures of 300 and 650 K and in 4000 pulses for a temperature of 700 K. For temperatures higher than 700 K, the optical absorption rapidly reaches an asymptotic value which is well below 1%. For the prototype facility the optical absorption is higher than 5% after 850 pulses for 750 K and after 1300 pulses for 800 K. The optical absorption saturates below 5% for 825 K. And for temperatures higher than 850 K, the absorption is below 1%.

We can assume that 5% is lens replacement limit. Lenses with such a reduced optical transmission will not fulfill the optical system requirements. Lens replacement or lens healing, e.g., by annealing [28] must be carried out. On the other hand, we can assume that 1% is the optimum operation limit, i.e., lens performance beyond this limit may be compromise the power plant operation. These results illustrate that: (i) for the experimental facility the lens replacement limit is reached in 2000 pulses, i.e., the lifetime of silica final lenses operating at room temperature is comparable to the expected facility lifetime. In addition, increasing the lens temperature above ~700 K would lead to a reduction of the optical absorption below the optimum operation limit. (ii) For the HiPER prototype facility, an optical absorption as low as 0.1% is estimated at steady-state operation temperature (938 K), i.e., much lower than the optimum operation limit. Moreover, under this scenario the optical absorption is observed to be over the optimum operation limit only if the temperature is lower than ~850K. (iii) If the steady state temperature in the HiPER demo facility could be controlled below the maximum service temperature, e.g., with an external cooler at ~900 K, the optical absorption would be lower than 0.5%.

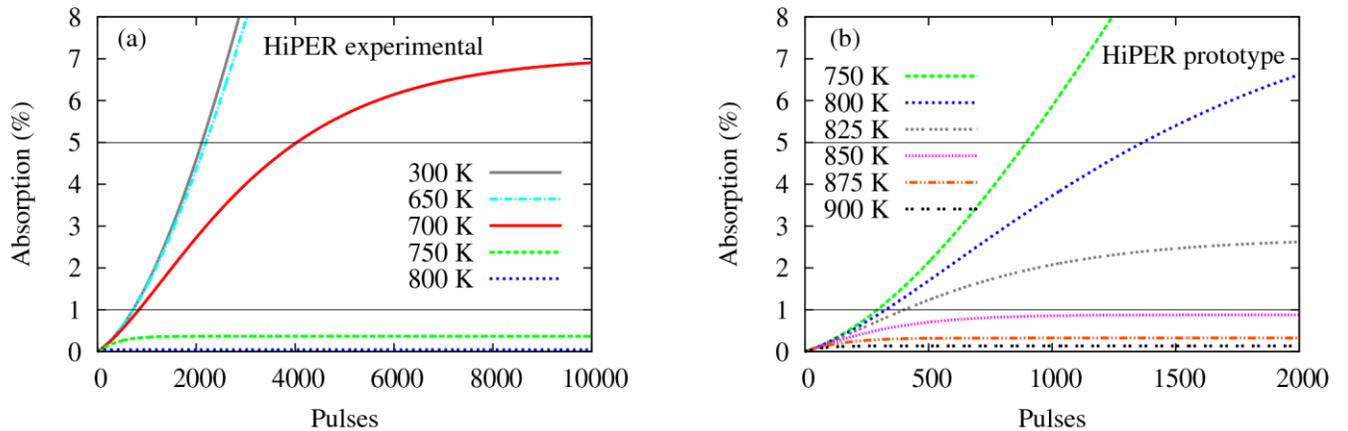

**Figure 9.** Optical absorption of a 5 cm thick silica lens at $\lambda=350$ nm for different temperatures with a 20 MJ yield target and a rep. rate of 0.5 Hz in the experimental facility (a), and in the prototype facility with a 50 MJ yield target and a rep. rate of 10 Hz (b).

The absorption as a function of the pulse number calculated at different temperatures for the LIFE.2 facility is shown in figure 10. Because of the small thickness of the lenses, the optical absorption for LIFE.2 increases slower than in the case of HiPER facilities. In steady state the maximum temperature reaches 744 K. Then, according to our calculations, the optical absorption is 3.9%, close to the reported value of 3.5% [3]. At temperatures below 744 K, the optical absorption is above the lens replacement limit (5%) in about 20000 pulses at 650 K and in about 23000 pulses at 700 K. The absorption remains below the optimum operation limit (1%) for temperatures higher than 775 K. Therefore, LIFE.2 lenses will present an optical absorption below the lens replacement limit in steady state operation and it can be reduced even further (below the optimum operation limit) if higher lens temperatures are reached, e.g., by means of an external heater.

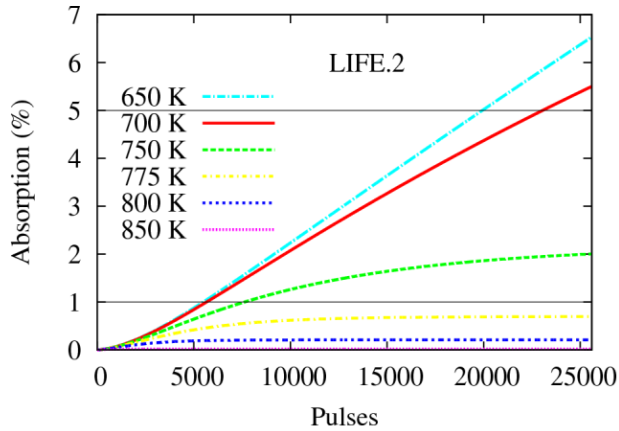

**Figure 10.** Optical absorption of a LIFE.2 final lens at $\lambda$=350 nm for different temperatures with a 132 MJ yield target and a rep. rate of 16 Hz.

Finally, the design points for HiPER (prototype and demo) and LIFE.2 facilities are depicted in figure 11. The colour map stands for the lens absorption. The design points for HiPER prototype and demo facilities, when keeping somehow the operation temperature at ~ 950 K, are located in the low absorption region, being lower than the optimum operation limit (0.1% and 0.5% for prototype and demo respectively). Two operation points for LIFE.2 are shown: (i) the so called beam heating which corresponds to the case calculated in this section ($A$ = 3.9%) where the laser beam heating effect on the lenses is considered, (ii) external heater in which the lens temperature is assumed to be somehow increased above 850K by an external heating source to reduce the optical absorption down to values of the order of 0.1% below the optimum operation limit.

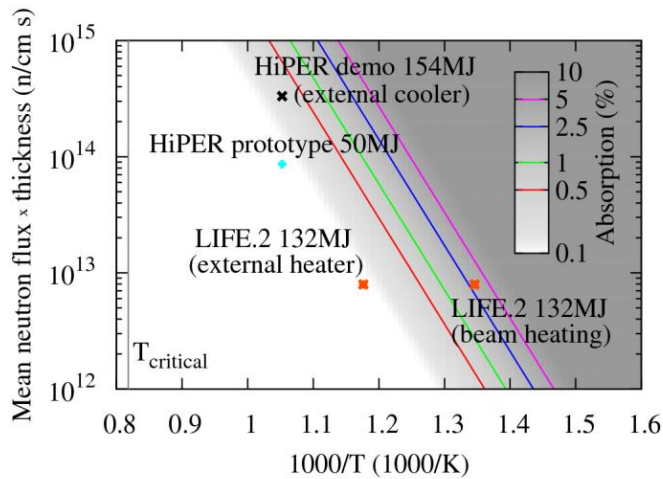

**Figure 11.** Parametric representation of the different HiPER and LIFE.2 facilities. The colour map indicates the absorption, $x$ axis is the inverse of lens temperature and $y$ axis is the mean neutron flux (by primary neutrons) multiplied by the lens thickness.

## 6. Reactor startup and optical restrictions

In the previous section, the strong dependence of the radiation-induced changes in the optical properties of the lenses has been illustrated. This dependence turns out to be a critical point during the reactor startup process in which operation temperatures are low.

To study the lens behavior during the reactor startup, the temperature (along $z$) and the colour centre concentration are calculated by means of a simplified 1D model. For these calculations the neutron and gamma dose rate, the neutron flux, as well as the silica properties (mean free path and number of defects formed following a primary knock on atom) are considered to be constant. The deposited laser energy is calculated every time step taken into account the absorption due to the concentration of colour centres.

The time evolution of the lens absorption during the startup process for the different facilities is shown in figure 12. For the prototype facility, an absorption as high as 17% is estimated during the first 400 seconds. After this time the optical absorption rapidly decreases being ~2% for times longer than 1200 seconds. This behavior can be explained by considering that during the first seconds a large number of colour centres are formed. In principle, one should expect the colour centre density to become larger when increasing time (increasing the number of shots), but his only holds if the lens temperature is low enough to prevent colour centre annealing. When the lens temperature is above a certain limit for colour centre annihilation to happen, the average colour centre concentration at any given moment depends on its formation and annihilation rate. Thus, the data shown in figure 12 illustrate that for times longer than 400 seconds the colour centre annihilation rate is higher than the formation one and therefore, the lens absorption decreases. The fact that the absorption coefficient during the first states of the startup process is significantly higher than the lens replacement limit (5%) might impede the startup of the reactor. A possible solution to avoid such an absorption enhancement may be to use an external heater to increase somehow the lens temperature during the startup process. As shown in figure 12, external heating of the lenses up to a temperature of 850 K reduces the optical absorption to acceptable limits. After the initial moments (around 2000 seconds) the absorption coefficient asymptotically decreases to reach the steady-state value below the optimum operational limit (1%). Note that the absorption calculated during the startup does not saturate to the values previously calculated in continuous mode due to the assumptions of the simplified 1D model. However, this does not affect the main conclusion, i.e., that unless we make use of an external heater the HiPER lens absorption will become unacceptably high during startup.

A very different time evolution of the optical absorption is observed for the Fresnel lenses of LIFE.2. In general, absorption increases smoothly, even during the initial stages, being below the lens replacement limit 5% at any given time, see figure 12. In view of these results, the Fresnel lenses would work properly in the HiPER demo facility. Figure 12 show a simulation of the optical absorption evolution exchanging HiPER lens for a Fresnel lens in HiPER demo reactor with its operational mode. Calculations demonstrate that a reduction of the lens thickness from 5 to 0.5 cm would prevent the absorption enhancement during the startup process, keeping the optical absorption below 2% for all times.

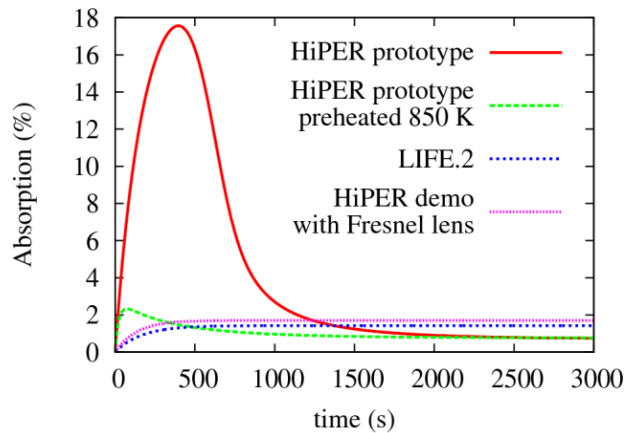

**Figure 12.** Calculated lens absorption evolution during the startup process for HiPER (prototype and demo) and LIFE.2 facilities. For the calculations the surrounding temperature is assumed to be constant and equal to 600 K.

*6.1. Focal length and aberrations in the HiPER lenses*
We need to address the dependence of the refractive index on temperature because a variation in the refractive index affects the focal length of the lens and therefore, the ignition process itself. The focal length is calculated by means of the thick lens equation in vacuum

$$\frac{1}{f} = (n_s - 1)\left(\frac{1}{r_1} - \frac{1}{r_2}\right) + \frac{d(n_s - 1)^2}{n_s r_1 r_2}, \qquad (5)$$

where $n_s$ is the refractive index, $r_1$ is the radius of curvature of the lens surface closest to the light source and $r_2$ the radius of curvature of the lens surface farthest to the light source. The focal length is one half the distance between the lens and the chamber center, i.e., $f = 4.0$ m in HiPER facilities. In addition, we consider that HiPER lenses are biconvex with a focal length of 4 m at 600 K and $r_1 = - r_2 = 382.3$ cm. In the worst case in the HiPER prototype facility, the lens temperature increases from 600 K, the surrounding temperature, to the steady state temperature 938 K, therefore, the refraction index increases from 1.47888 to 1.48346 (see table 5). If we assume that the lens geometry is fixed ($r_1$ and $r_2$ constants) and interpolate the refraction index for the experimental data in reference [41], we obtain a variation of 3.8 cm in the focal length, This deviation is unacceptable and requires a correction method to allow for accurately focusing of the laser. The lens system must be designed to keep the focal length under control. The most appropriate solution is to use an external heater to keep constant the lens temperature before and during the startup.

**Table 5.** Refraction index for the indicated temperatures (from Ref. [41]) and the resulting focal length for biconvex HiPER lenses.

| Temperature (K) | Refraction index at 350 nm | Focal length (cm) |
|---|---|---|
| 600 | 1.47888 | 400 |
| 760 | 1.48110 | 398.2 |
| 938 | 1.48346 | 396.2 |

The temperature profile along the $z$ axis presents a variation of about 100 degrees in steady state operation, see figure 5(a). This leads to the appearance of a small aberration effect. Along the radial direction, the temperature varies from 938 K in the centre to 760 K at $r=30$ cm (corresponding to the laser edge). This temperature profile introduces a longitudinal aberration of 2 cm at 8 m (the chamber centre) this corresponds to a transversal aberration of 75 μm. A way to mitigate this aberration is to allow neutrons reach an area larger than the laser area to produce a smoother radial temperature profile.

Another important feature is that the temperature profile reached at steady state solution into the lens of the HiPER prototype facility induces a surface deformation. Small deformations move the focal length beyond the target point. In the prototype facility, the lateral surface has an axial displacement of 30 μm. Assuming a spherical lens, the sagitta of the spherical surface can be obtained by $s = r - \sqrt{r^2 - (D/2)^2}$, where $r$ is the radius of the spherical surface and $D = 0.75$ m is the diameter of the lens. After the surface deformation, the final radius of the spherical surface $r_f$ can be calculated taken into account the sagitta variation $\Delta s = 30$ μm,

$$r_f = \frac{(D/2)^2 + (s \pm \Delta s)^2}{2s}. \qquad (6)$$

We have calculated the focal length of the deformed lens with the thick lens equation in vacuum (5). This curvature variation is feasible for biconvex thick lens ($r_1 = -r_2$) with 4 m focal length because this distance varies around 50 nm. Although, converge meniscus thick lens, with 4 m focus length and $r_2 = 8$ m to reduce laser reflexion, has worse performance, its focal length varies up to 0.1 mm. This deviation can be corrected keeping the same temperature at the inner and outer surfaces of the lens. This solution implies a complex system for each lens in a reduced space. Although the optical restrictions induced by the thermo-mechanical response in HiPER lenses can be corrected with external systems, a new scheme of the final optical components must be studied for future power plants.

## 8. Conclusions

A systematic study on the response of HiPER and LIFE final lenses to realistic irradiation conditions is presented in this paper. HiPER experimental (20 MJ explosions in bunch mode), prototype (50 MJ explosions in continuous mode) and demo (154 MJ explosions in continuous mode) scenarios as well as LIFE.2 (132 MJ explosions in continuous mode) scenario are considered to calculate the radiation fluxes. LIFE will employ indirect targets. Their explosion generates a prompt X-ray pulse so intense that no material can withstand it. As a radiation mitigation strategy LIFE has adopted the use of chambers filled with residual gas. Therefore, from the point of view of the final lenses the major irradiation threads are related to the arrival of neutron and gamma pulses. On the other hand, HiPER will employ direct targets, which generate an intense ion pulse that target facing materials must withstand. In this case, residual gas mitigation strategies cannot be adopted due to incompatibilities with the target injection. However, thermo-mechanical calculations show that the ion pulses in every HiPER scenario lead to a final lens temperature above the silica melting temperature. Therefore, some sort of mitigation strategy, not described in this paper, will be necessary to get rid of the ion irradiation of the final lenses.

Disregarding ions, the remaining X-ray, neutron and gamma irradiation is observed to increase the HiPER lens temperature after every shot. The estimated temperatures in steady-state conditions are below the silica melting temperature for HiPER experimental and prototype facilities and above the melting point for the HiPER demo facility. This must be a major consideration for the design of the HiPER demo reactor optical system. In the case of LIFE.2 the position of the final lenses in addition to the residual gas make possible to maintain the lens temperature below the melting point.

The temperature profiles along the $z$ and $r$ directions drive to stress generation. In all cases the calculated stresses are observed to be lower than the silica tensile strength (48 MPa), which indicates that, silica lenses can withstand the radiation-induced mechanical stresses. The critical crack length for brittle fracture is calculated to be 6.6 mm, 2.04 mm for HiPER prototype and demo facilities, respectively. The lenses in HiPER experimental and LIFE.2 facilities will not suffer from brittle fracture since the estimated critical crack length is larger than the dimensions of the maximum traction volume of the lens. The estimated stress intensity factor range is significantly smaller than the silica fatigue threshold. This indicates that negligible crack growth due to cyclic traction stress will take place in any facility.

Particle and purely ionizing radiation generate colour centres. The concentration of colour centres (oxygen deficient, ODC, and E´ centres) at a given moment depends on their formation/annihilation ratio, which ultimately depends on temperature. The optical absorption at 350 nm (laser wavelength) depends on colour centre concentration. In steady-state operation, the optical absorption is calculated to be lower than the optimum operation limit (1%) for the different HiPER reactors. In the case of LIFE.2, this also occurs provided that an external heater keeps the lenses above 775 K.

We also discuss the lens performance during HiPER reactor startup. We show that before reaching steady state conditions the optical absorptions is unacceptably high, it exceeds the lens replacement limit (5%) because the lenses are too cold. Some strategies must be used, for example, the use of external heating to pre-heat the lenses at temperatures of about at least 850 K. Alternatively, Fresnel lenses (as in LIFE) appear to be less prone to excessive optical absorption during startup. The focal length variation during startup is another important issue. The temperature variation during startup leads to changes in the refractive index and therefore in the focal length that turn out to be unacceptable. Once again the proper use of an external heater could be used to keep the lens at a constant temperature during the startup procedure. Once steady state is reached, temperature profiles along both axis reached may produce aberrations. A way to make the temperature profiles as smooth as possible is to irradiate an area larger than the laser area or reduce the lens thickness (Fresnel lens option). In summary, we have addressed a number of fundamental and technological problems related to the performance of silica-based lenses in LIFE and HiPER laser fusion facilities. Some problems have been identified and require further efforts to find optimum solutions.

## 9. Acknowledgments
The authors thank the Spanish Ministry of Science and Innovation for economical support via the ACI-PROMOCIONA program 2009 (ACI2009-1040).